\documentstyle[12pt]{article}

\begin{document} 

\begin{center}  
{\large {\bf  One World, One Reality, and the Everett Relative State Interpretation of Quantum Mechanics}}
\vskip.75cm
 {\bf  Paul L. Csonka}
\vskip.5cm
 
\end{center}

 \begin{abstract}
 We modify the Relative State Interpretation (RSI) of Quantum Mechanics so that it does not imply many worlds and parallel realities. We drop the assumption that probability amplitudes correspond one-to-one with reality: Not all information is contained in amplitudes, and not all amplitudes need be realized. Amplitudes do not ``collapse'' after a measurement, but evolve continuously, including unrealized ones. After each ``event" only one  possible outcome is realized. Therefore, if a value is measured, that value is real, all others are not; there is only one reality and one world. Reality content is ``quantized" : unity for realized outcomes, zero for all others. It is ``conserved'': can move along any possible sequence of events, but only one at a time. The modified RSI is is strictly deterministic in the sense that the ``global'' probability amplitudes of the universe are determined for all times by the laws of physics and the initial conditions. They guide all events. But it is not deterministic in the sense that from the amplitudes one can not predict which outcome actually happens; that represents new information that accumulates as history unfolds. To the extent that  information is part of the physical world, the coming into being of the universe is ongoing, even after the Big Bang. All predictions of the two versions agree, except possibly in esoteric, untested cases.
\end{abstract}


\section{ Introduction}
 
Quantum Mechanics provides a prescription for calculating the outcomes of physical experiments. It has never been contradicted by observation. On the other hand, the interpretation of these calculational rules proved to be difficult.	
	
One may legitimately claim, as I sometimes do, that the rules themselves suffice, and no deeper understanding is needed. \cite{Lu, Pe, Be} However, in that case an important source of intuition is lost  \cite{Ma},  a source that may have helped in the understanding of how those very rules could be modified, or how to apply them in unfamiliar circumstances. Therefore, efforts to find alternative interpretations continue to be of interest.
		
The traditional ``Copenhagen'' Interpretation (CI) holds \cite{NB, vNeu} that the wave function contains all known information about the system, and determines the probability of all measurement outcomes.  It changes continuously according to some differential equation, but whenever a new measurement is performed, the wave function undergoes a discontinuous  change (``collapses'') instantaneously and everywhere, to adopt itself to the new information provided by that measurement. While this interpretation agrees with all observations, it is incomplete at best. Partly because of difficulties associated with any instantaneous change, and mostly because there is no clear definition of what constitutes a ``measurement''. Postulating that any observation performed by an ``entity with consciousness'' is a measurement \cite{Wi}  is insufficient, because  ``consciousness'' itself lacks precise definition.

An alternative, fundamentally different and very attractive interpretation is the ``Relative State'' Interpretation (RSI), also known as  the ``Many Worlds'' Interpretation due to H. Everett. \cite{Ev} Its implications were further elucidated by J.A. Wheeler, B. De Witt, and others. \cite{Whe, DeW, dEs}  According to it, too, the wave function determines the probability of all events, but never undergoes any abrupt ``collapse''. It changes continuously everywhere at all times, whether or not a measurement has been performed. All possible outcomes of a measurement are simultaneously realized, and are equally real, indeed, the amplitude is assumed to be in one-to-one correspondence with reality. This leads to parallel realities, perhaps infinitely many of them, and a corresponding multiplicity of parallel worlds. This interpretation not only fits all observations, but it is also  complete and self consistent. Indeed, when supplemented with insights concerning the nature of coherence, decoherence, and effects due to the physical environment,  \cite{Ze, Ge, Ha},  the ``Relative State'' or ``Many Worlds" Interpretation seems to be essentially the only complete and fully self consistent interpretation Quantum Mechanics available today. 

In addition to these two interpretations, the literature contains several other ingenious proposals how Quantum Mechanics, or various aspects of its, should be understood. For the following argument, however, it will be adequate to restrict our discussion to the two interpretations just mentioned.

Despite the completeness and self consistency of  the Relative State Interpretation, many find it disturbing to assume the existence of innumerable simultaneous realities and worlds. They feel that this violates the expectation that science present the simplest interpretation of events, by postulating the existence of who knows how many worlds, and even worse, most of them undetectable in principle. It also seems thereby to have given up on the fundamental goal of the natural sciences: to know the universe. 

The question arises whether one could modify the Relative State Interpretation so that the modified version remain complete and self consistent, but without the assumption of many worlds and parallel realities. In the following I will argue that such is the case. Since that possibility does not seem to be generally appreciated, it will be briefly discussed here. 

The modified version of the Relative State Interpretation to be explored, leads to the same experimental predictions as the standard version, except perhaps in rather esoteric situations to be mentioned below. This writing does not argue in favor of either version. Its purpose is to demonstrate that based on what is known today, either may legitimately be adopted.

\section{ Modified  ``Relative State'' Interpretation}

In the standard version of the Relative State Interpretation it is assumed that the probability amplitudes are in one-to-one correspondence with reality, in effect the amplitudes {\it are} reality. Based on the mathematical formalism alone, this is the most straightforward, and therefore a very natural assumption to make. However, letting oneself be guided by formalism alone is not necessarily optimal. Indeed, unless one properly delimits their domain of applicability, formalisms in their full generality, including symmetries and equations, can lead to manifestly unphysical results.  

In the modified version we will drop the assumption just mentioned. Specifically:  (1.) The modified version does {\it not }assume that all event outcomes with non-zero probability must actually occur in the world. (2.) It does {\it not} assume that the calculated probability amplitudes contain all information about the world. In fact, as will be discussed below, according to the modified version: 

    (1.) If a measurement indicates that a certain quantity  x  has the value, say  $x_{1}$,  then it is understood that in the real world the value of that quantity is $x_{1}$, this is the {\it only} reality. The other possible outcomes of the measurement event have not been realized, there are {\it no} parallel realities, {\it nor} parallel worlds, even if there are other outcomes whose probabilities are non-zero. This understanding is possible, because it is not assumed that all event outcomes with non-zero probability must be realized.
 
    (2.) When a measurement is performed, then although the available information about the world changes (since that measurement supplied a previously unknown value, say $x_{1}$), nevertheless the originally calculated probability amplitudes do not ``collapse'' or get altered in any way as a result of the experiment. Even though there is only one reality, all amplitudes continue to develop smoothly, including the amplitudes of those outcomes that were not realized, and all remain normalized to the original initial state. It is possible to postulate this, because the amplitudes are not assumed to contain all information about the world; in particular, they do not contain the newly found experimental result, only its probability.
  
To state more precisely the difference between the standard and revised versions of the RSI, we first need to deal with terminology.

Consider a physical system  $S$, composed of two objects  $A$  and  $B$.  After an interaction between $A$  and  $B$,  the state of  $S$  can in general be written as 
\begin{eqnarray}
\chi = \Sigma_{j,k}a_{j,k}\psi_j\varphi_k
\label{1}
\end{eqnarray}
where \{ $\psi_{j}$\} and \{$ \varphi_{k}$\} are  complete sets of orthonormal basis states referring to  A  and  B  respectively, and the  $a_{j,k}$  are coefficients. (If the  $j$ or $k$ assume non-discrete values, the  $\Sigma_{j,k}$ is to be understood as an integral over those.) Each of the $\psi_{j}\varphi_{k} $ for which  $a_{j,k} \not= 0$, describes one of the  states in which the (A+B) system may be found after the interaction. We will refer to any of those states as a possible `` outcome'' of the interaction in question.  We mean by a ``branch"  of the state  $\chi$ an amplitude described by any one term in the sum in Eq.(\ref{1}).  

If  one can find coefficients $b_{j}$ and $c_{k}$ such that  $a_{j,k} = b_{j} . c_{k}$  for all $j$ and $k$,   then $\chi$ can be expressed in a simpler form, as  $\chi =  \alpha . \psi . \varphi$, with $\alpha$ a coefficient, and $\psi$ and $\varphi$ being some general states (not necessarily one of the basis states mentioned above)  referring to system A and B, respectively. In such a case the interaction under discussion has only one possible outcome. However, if no such $a_{j}$ and $b_{k}$ can be found, then the sum in Eq.(\ref{1}) has always more than one terms, no matter which set of basis states \{ $\psi_{j}$\} and \{$ \varphi_{k}$\}  are chosen. Then $\chi$ describes what is referred to as an ``entangeld" state of the system. Whenever the final state is``entangled", there are more than one possible outcomes. 

In the following, an ``event'' will mean an interaction between physical systems for which the resulting amplitude describes an entangled state that differs from what the result would have been be without the interaction (in addititon to perhaps an overall change in phase). For example, the scattering of two particles on each other is an ``event'', if the total outgoing amplitude differs from what it would have been without the interaction, so that there is a finite probability that the particles  would either experience a momentum change as a result of the scattering, or else continue without any change. It is an event, because the final state differs from the initial state, and there are at least two possible outcomes. It is an ``event" even if both particles are in fact observed to continue moving without momentum change, since it is the ${\it amplitude}$ change that counts, and not what actually happened. On the other hand, a particle in the ground state of a fixed potential well does interact with the potential, but this interaction leads to only one outcome: it leaves the ground state unchanged \footnote{Except (in the Schroedinger picture) for an overall change in phase. }, therefore it is not an ``event''. 

A ``measurement'' is an event in which one of the interacting systems is a measuring apparatus, as understood according to the Relative State Interpretation.

We recall that once the initial state of any physical system and the laws that govern it are given, one can calculate the probability amplitude for each possible state into which the system can develop. This can be done for example by using the Feynman path integral method, in which one adds up contributions from all possible pathways leading to the outcome of interest. The result depends on the contributions of all pathways that the system {\it could} follow, but is independent of which pathway the physical system actually {\it does} follow.

In principle such a calculation (if performed immediately after the appearance of the universe as we know it) could evaluate for all times a ``global probability amplitude'' containing the amplitudes of every outcome of every subsequent event that can ever occur in the world. As time goes on, more interactions occur, and there are increasingly many possible event outcomes. Therefore, the probability of any one outcome tends to decrease with time. Since however the total probability of all the outcomes at any one time remains unity, the sum of all ``global'' probabilities always remains normalized to the original initial condition. \footnote{Of course it may be convenient to change normalization for purposes of a calculation.} Such is assumed to be the case in both the standard and the modified versions of the Relative State Interpretation. The difference between the versions is in how one understands what the amplitudes mean.

In the standard version it is postulated that (1.) All branches of an entangled state are realized, and form different ``realities", or ``worlds".  (2.) All amplitudes at all times continue to develop according to the relevant differential equations. 

In the modified version, on the other hand, one assumes 

{\bf Postulate 1 : For every event there is at least one set of  \{$\psi_{j}$\}   and \{$\varphi_{k}$\}  with the property that when one writes the outgoing state  $\chi$  produced by that event in terms of \{$\psi_{j}$\}   and \{$\varphi_{k}$\} as in Eq. \ref{1}, one and only one term in the sum represents an event outcome that is realized.  
\footnote{If either  \{$\psi_{j}$\} or \{$\varphi_{k}$\} contain more than two states, then there must be infinitely many such sets (in addition to physically uninteresting changes of phase for the basis states), since any unitary transformation that leaves the realized $\psi$ and $\varphi$ unchanged, leads to another set with the same property.} 
All other terms represent unrealized outcomes. }

{\bf Postulate 2 :  All amplitudes, whether realized or not, continue to develop according to the appropriate differential equations. }  

The outcome realized after an event, remains realized until the system encounters the next event, at which point the new outgoing state, too, will have one of its outcomes realized, while all others will not be. Loosely speaking, at every event reality "decides" which branch to follow, but always follows only one branch. Which branch that will be can not be predicted, but since one and only one branch of a physical system is realized, there is always only one reality, and one real world. 

By the same token, if A (or B) itself is composed of two physical subsystems, only one branch of {\it its} state can be realized at any given time; and if a sub-system itself contains constituent systems, only one branch of  any constituent system's state can be realized, and so on. Consequently, if one knows that  $S$ is composed of  $N$  basic constituents, then its state $\chi$ can be written in analogy with Eq. \ref{1}, as 
\begin{eqnarray}
\chi = \Sigma_{j_{1}, j_{2}, ... , j_{N}} a_{j_{1}, j_{2}, ... , j_{N}}  \psi_{j_{1}}\varphi_{j_{2}} ...  \xi_{j_{N}} ,
\label{1a}
\end{eqnarray}
where the states $\psi_{j_{1}}\varphi_{j_{2}} ...  \xi_{j_{N}}$ refer to the $N$ constituents, and we know that at any given time only one of the terms in the sum will be realized, while all others can not be.

In practice it may happen that there is a set,  $G$,  of   $N_{G}$   constituents  with the property that the correlations between them and all other constituents of  $S$  do not depend (or depend only negligibly) on the state of all the individual members of  $G$, but only on some aggregate property of them, such as the center of mass coordinates of $G$, or its total angular  momentum. In that case it is often convenient to consider  the entire $G$  as a constituent object of  $S$, with the appropriate center of mass, angular momentum, and possibly other defining parameters, and refer to the remaining characteristics specifying the exact state of the $N_{G}$ constituents as describing the ``internal state" of
  $G$.

Although one can not foretell which of the possibly many outcomes will be realized, nor in which set of basis vectors that term would be easiest to write down, one can calculate the probability that a particular outcome will be realized. What is that probability? It is precisely the quantum mechanical probability of the outcome in question, as calculated from the amplitudes themselves. \footnote{This is simply the consequence of the fact that the quantum mechanical probability gives the likelihood that a certain outcome is observed to exist, and that if an outcome is observed to exist then it is real.} 

When referring to realized v.s. unrealized states, it is convenient to adopt a language similar to that used when talking about electric charges: Consider a collection of neutrons  and atomic nuclei (all objects that can be either electrically neutral, or carry a positive but not a negative charge).  For this system, when we say  that the system is electrically neutral, we mean that none of the particles carries a charge, i.e. they must all be neutrons. On the other hand, if we say that the system is positively charged, we do not mean to imply that every particle in the system is positively charged, we only mean that the total system carries some positive charge, i.e. it must contain at least one nucleus.  Furthermore, looking more closely, we may observe that the positively charged nucleus is itself is composed of constituent particles, namely protons and neutrons. If we then say that the nucleus itself is positively charged, we do not imply  that all particles within it carry a positive charge, only that it must contain at least one proton. Similarly, if we say that an entangled state is not realized, we mean that none of its branches are realized. On the other hand, by saying that an entangled state is realized, we only mean that one of its branches is realized. Furthermore, looking more closely, one may observe that the realized branch itself is an entangled state of constituent systems. That implies that one of {\it its} branches is realized, and does not mean that all of them are. And if that realized branch itself is also entangled, then one of {\it its} branches must be realized, but not the others, and so on.

According to the CI,  probability amplitudes change (``collapse'') as determined by the result of experiments, therefore amplitudes depend on what occured earlier. According to the standard version of the RSI, on the other hand, amplitudes are not changed by the outcome of experiments, and it does not even make sense to ask which possible outcome happened earlier, since all of them are assumed to have happened. By contrast, according to the modified  version it certainly does make sense to ask which possible outcome did in fact occur, yet all probability amplitudes remain unaffected by which outcome of an experiment has been realized. There is no inconsistency here, because the amplitudes, as given e.g. by the Feynman path integral method, depend on the contributions of all {\it possible} outcomes, and not on whether all outcomes, or some outcomes, or which outcomes {\it actually} occured. 

According to both versions of the RSI, as soon as the initial conditions and the laws of physics are given, the global probability amplitudes are determined for ever. According to the standard version, they represent all the various parallel realities into which the universe will evolve. On the other hand, according to the modified interpretation, the pattern of amplitudes constitutes merely a framework within which all events must unfold; they circumscribe the possible outcomes, but are unaffected by them. (The laws of nature also behave this way: They specify such things as the space-time metric, the equations of motion, which quantities are conserved, and so on, and thereby delimit what is allowed, but do not depend on what actually happens.) The outcome of an experiment does not affect the amplitudes, all amplitudes, including those that are not realized, continue to contribute their share to the value of subsequent probability amplitudes. The outcome influences subsequent events only by virtue of the fact that they specify the value of measured quantities, and thereby restrict the initial conditions from which later events evolve. By looking at probability amplitudes only, one can not tell which event outcome was actually realized, that question can be answered only by consulting the result of an observation of that event.

One can represent the real state of the entire universe at some time, t, by specifying the endpoint of its realized state vector in a very high dimensional Hilbert space, thus giving the values of a complete set of observables that characterize the universe at that time. As a result of each of the multitude of events that keep happening in the world, choices occur:  from the set of all possible event outcomes one particular set is realized, the state vector of the universe adjust itself to that realized set, and so the representative point in Hilbert space moves. As history evolves, the point describes a trajectory depending on the realized history of the universe. Which possible event sequence will in fact be realized, i.e. which trajectory the universe will in fact follow, can not be predicted from the global probability amplitudes, but the likelihood of each sequence is determined by them. 

 As a crude analogy, one could represent the probability amplitudes by a pattern of crisscrossing ever branching highways, some wider, others narrower; and symbolize reality by a car traveling along these highways.The highway pattern defines all pathways along which vehicles can move, more easily along some, less so along others, analogously to how the amplitudes circumscribe the possible outcomes of events, some more probable, others less so. The exact route along which any given car will actually travel, however, represents additional information, not contained in the highway grid, just as the actual outcome of any event represents new information, not contained in the amplitudes. Furthermore, the fact that a car passed over a certain sequence of roads and arrived at a point  $P_{1}$  , does not change the layout of highways; it only specifies that the next portion of the car's route has to start at  $P_{1}$ ; similarly, the fact that a certain outcome was realized does not alter the pattern of amplitudes, only specifies the initial conditions from which the next sequence of events has to emerge. 
\footnote{In the classical limit only a single pathway remains, its probability is one.}

\section{ Measuring instruments}

Observations are designed to measure certain ones of a complete set of eigenvalues, $\alpha_{j}$, characterizing the object. Therefore it is convenient to choose the  $\psi_{j}$\  states referring to  the object, to be eigenstates  belonging to the eigenvalues to be measured.  The $\varphi_{k}$ states describing the observing instrument, on the other hand, depend on the design of the instrument, possibly a human. A macroscopic instrument  (including all its parts, such as the recording equipment) is composed of many particles. In such cases the $\varphi_{k}$ describe states of a macroscopic system that may have many more degrees of freedom than the object under observation. Some of these may be coordinates (e.g. the position of the tip of the pointer) that specify the ``macrostate" of the instrument that determines the instrument reading. The remaining degrees freedom may represent various internal coordinates (e.g. the positions of atoms within the pointer) that specify the  ``microstates" of the instrument, each microstate characterized by eigenvalues of an appropriate set of commuting observables. Therefore, a macroscopic instrument in a given macrostate can be in any one of a large number of microstates. It is usual to refer to all these microstates as ``belonging" to that macrostate.
One can rewrite Eq.(\ref{1}) as 
\begin{eqnarray}
\chi = \Sigma_{j,k}a_{j,k}\psi_{j} \varphi_{k} =\Sigma_{k}\alpha_{k}\xi_{k}\varphi_{k}
\label{2}
\end{eqnarray}
Here $\alpha_{k}\beta_{j,k}=a_{j,k}$, and the $\xi_{k}=\Sigma_{j}\beta_{j,k}\psi_j$ states refer to the object. The $\xi_{k}$ can be normalized to unity, but generally will not be  mutually orthogonal. 

A perfect instrument has to satisfy three requirements:  (1.) It has to be completely ``faithful", so that when $\varphi_{j}$ is in a given macrostate, the instrument reading is perfectly correlated with one and only one state of the object.  (2.) To avoid ambiguity, (a.) the  \{$\varphi_{k}$\} macrostates have to be ``clearly distinguishable" from each other, and (b.) each \{$\varphi_{k}$\} has to be correlated to  clearly distinguishable \{$\psi_{j}$\} states. (3.) the instrument should have ``perfect memory", to ensure that what was observed will not be forgotten. In practice no instrument is ideal, but to be useful it has to satisfy these conditions at least to a good approximation. To the extent that the instrument satisfies conditions (2.a.) and (3.), then as a result of interactions with the surroundings, soon after the measurement these macrostates decohere, linear superpositions of them will no longer be observable. 

If the instrument is perfectly faithful, then all microstates $\varphi$ belonging to the same instrument macrostate $\varphi_{k}$ will be correlated with only one $\xi_{k}$. If also the instrument readings are  clearly distinguishable, then the $\xi_{k}$ are mutually orthogonal. Ideally one aims for $\xi_{k}=\psi_{k}$, in which case the last requirement is satisfied. When we see the instrument in the state $\varphi_{k}$ after a measurement event, we know that according to the modified interpretation this $\varphi_{k}$ macrostate is real, the other $\varphi$  macrostates are not, and that the object is in state $\xi_{k}$, and no other state. We do not know which microstate is realized, but measuring additional suitably commuting observables one can find out more details about the realized state.  If any one of the particles of the realized state is itself composed of constituent particles, then only one term in the sum describing $\it{their}$ entangled state is realized, the others (if any) are not, and so on. 

If a certain outcome of an experiment is the real one, then of course no subsequent experiment can contradict that fact. If a clearly distinguishable and lasting record of this result is made, then the amplitudes of the various outcomes can no longer be coherent with each other, and so the observable predictions of the modified version agree with those of the standard version (and also with those of the CI ). On the other hand, if the various outcomes remain coherent, or become coherent again, then they may subsequently interfere with each other. The predictions of the two versions are nevertheless  still in agreement (but not necessarily with those of the CI ), since the probability amplitudes do not depend on which outcome was the realized one. According to the modified version, however, we can assert that only one of the outcomes has been realized.

\section{ Examples}

The simplest way to illustrate various aspects of the revised RSI, is to consider  some much discussed,  one might say notorious, thought experiments.

 {\bf Electron crossing a grating}: Given the initial conditions, one can calculate the probability amplitudes for finding the electron at various points behind the grating. When it is observed that the electron has actually impinged on a detector surface at point $x_{1}$, the electronÕs position becomes known. This represents new information. Nevertheless, according to the modified version, just as in the original standard RSI, no ``collapse'' of the wave function is assumed, and no renormalization of it is performed. The probability amplitudes continue to develop smoothly according to the guiding differential equation, and continue to mean what they have always meant: they determine the probability of finding the electron at various points behind the grating, for the originally given initial conditions. The knowledge that the electron actually arrived at $x_{1}$, is information not contained in the probability amplitudes, and allows one to conclude that there are no real worlds in which the electron arrived at some other point $x_{2}$,  or $x_{3}$, ...  .  No parallel realities or parallel worlds are assumed. What we see is what there is, and reality is unique.

{\bf Radioactive decay} : Radioactive material is placed near a counter which is switched on during some time interval. From the initial conditions one can calculate the probabilities that  the counter will detect exactly 0, or 1, or 2, or ...   , particles. When a measurement shows that in fact, say,  8 particles have been detected, this knowledge represents new information, not contained in the previously calculated amplitudes. Nevertheless, no ÒcollapseÓ and no renormalization of the wave function occurs. Furthermore, this experimental result tells us that in the real world events followed a course that caused the detection of exactly 8 particles.That could have happened in several ways, depending on which of the particles have decayed, which decay products reached the detector, and so on. We do not know which of these possibilities actually occurred, but they must have been such that they caused the detection of 8 particles. We conclude that any sequence of events that would have caused the detection of a different number of particles, is not real; a world in which one of those took place cannot be the real world.

If simultaneously other measurements are also performed, then additional information can be collected, such as the direction of travel of certain decay products. We will then know that any outcome not compatible with  {\it everything}  that has been measured, can not be real. There is still only one world and one reality.   

{\bf EPR-type Correlation experiments }: We observe that a particle has decayed, and know that the only decay channel is into two particles,  $a$  and  $b$,  both with  spin $1/2$, and that the two decay products must have total spin zero. The state of the $a$  and  $b$ particles is therefore entangled, and only one of the outcomes is real. We don't know which one that is, and the amplitudes of all outcomes contribute when the sate is measured. Suppose that at point   $P_{1}$  the spin z-component,  $s_{z}(a)$, of particle   $a$  is observed to be   $+ 1/2$. The probability amplitudes have not  ``collapsed'', but now we know that as a result of the measurement event particle  $a$ is in a state with    $s_{z}(a)= +1/2$   in the real world,  and that any other value is not real. In addition, since the total spin must vanish, we know that particle  $b$  must be in an eigenstate  of  $s_{z}(b) = - 1/2$. If at point $P_{2}$ that spin component is measured, the result will confirm this expectation. 

Suppose, however, that at $P_{2}$ one measures not $s_{z}(b)$ but  $s_{x}(b)$, and for it the value  $+ 1/2$   is found. Then we know that as an outcome of that measurement event the system realized the eigenstate in which $s_{x}(b)$  has that value, and no other eigenstate can be real. The probability amplitudes did not ÒcollapseÓ, but there is still only one reality, one world, and in it $s_{x}(b)= + 1/2$ .  
                                 
{\bf Stern- Gerlach experiment} : An electron with spin component   $s_{z}= +1/2$ along  z   is sent into a magnetic analyzer. Electrons with spin component  $s_{x} = +1/2$  along the x axis will be guided by a magnetic field along some path to the right, and those with  spin $s_{x} = -1/2$ along a path to the left. To the extent that during the interaction the analyzer does not change its state, the electron will continue in both spin states, behaving as it would in a fixed potential well. We know that if now the two paths of equal length are reunited, then once more an electron with  $s_{z} = +1/2$ appears. However, if there is a finite probability that the analyzer changes its state as a result of the interaction, then with that probability either the state with  $s_{x} = +1/2$,  or the one with $s_{x} = -1/2$ will be realized, but not both. In that case when the two paths  are reunited, only the realized state will be present, not the other. On the other hand, if the experiment is so cleverly arranged that following the interaction, after some time the analyzer and all of its surroundings end up in the same state for both spin states, i. e. the two spin states become coherent again (a feat difficult to accomplish), then reuniting the two branches will again result in an electron with  $s_{z} = +1/2$. Even though for a while only one of the two possible  $s_{x}$   values has been realized, both branches will now contribute to $s_{z} = +1/2$ outcome.  Since in such a case no record can survive that would indicate which the realized value was, that information is lost and can not be retrieved. Even so we know that according to the modified interpretation, for a time only one of the two states was real, not both.
                  
     {\bf SchroedingerÕs pet, and Wigner's friend} : A dog is placed in a closed box together with a  radioactive atom (assumed to be infinitely heavy, to simplify discussion), a detector, and a lever arm. At time  $t_{0}$  the atom decays, and a particle A is emitted. If the detector records A being emitted into the right hemisphere, the lever arm places food into the box. If A is detected in the left hemisphere, the  lever arm drops a ball into the box. From the initial conditions one can calculate the probability amplitudes for both sequence of events, and at the initial moment the emitted particle is in a state containing amplitudes for both outcomes. Subsequent to time  $t_{0}$  Wigner's friend looks into the box, records the result, informs Wigner, who in turn communicates the information to a wide audience, each of whom may record the received information, pass it on to others, and so on. 
     When does the choice between the two possible outcomes occur? According to the CI, as far as Wigner is concerned, the choice occurs when he learns the outcome; while as far as his friend is concerned, it occurs when {\it he} looks and and finds out. When does the choice  {\it really} occur? According to the CI the answer is not clear, because  ``observation" has not been defined. (Wigner sought to resolve this dilemma by arguing that the choice requires a ``conscious" observer, but left that concept undefined.) According to the standard RSI, this question makes no sense, because all outcomes happen. On the other hand, according to the modified version, a choice does indeed occur: once Wigner's friend has seen the dog play with the ball, it is clear that the second of the two sequences has taken place, and not the first. All records, including Wigner's memory, and the memories of the possibly thousands of people who knew the outcome, would serve as evidence. However, even before the observer looked, the dog already knew and started playing, therefore already then the second sequence was realized. Actually, even before that the lever arm has been activated, and even before that the emitted particle was detected in the left hemisphere, so that already by then only the second of the two possible outcomes was real. 
(And if on its way to the detector, particle A collided with some other particle B, resulting in  the particles changing their state, then even though we may have no record of that, after such an event only the second outcome was real.)

As in the previous examples, the wave function does not collapse, is not renormalized, reality remains unique.   

What if the detecting instrument detects not whether an atom has decayed or not, but observes in which  (possibly very unlikely) superposition of a decayed and an un-decayed state the atom is? In that case the outcome of the observation would tell us  in which superposition the atom is, and furthermore, that it is certainly not in any other superposition. The ``one world, one reality'' rule still applies.

{\bf Free v.s. interacting particles} :  

All components of the state describing a single particle moving freely, are equally real. The same is true for a particle moving in a a fixed potential well. For example, when the wave packet of a proton is passing, say, between fixed objects in a box, its state accommodates itself to the boundaries, including walls, and all values (or momentum components, etc.) in this wave packet are equally real. The whole state is just one vector pointing in some direction in Hilbert space, in fact, it may be chosen as one of the basis states \{$\psi_{j}$\} describing the proton.  Consequently, it does not make sense to ask questions such as : which component of the state "really" represents the proton's state, e.g. where exactly the proton "really" is. However, as soon as the proton affects any of those other objects such that their state becomes detectably different depending on e.g. where the proton is, then detectable  correlations appear, with it entanglement with significant probabilities, and  the difference between realized and non-realized branches of the combined amplitude has to be taken into account.

Of course no physical system is truly free, all systems  interact with background photons, gravitons, and what have you, and those may all produce events. But as long as  these interactions are weak enough,  and the system experiences no other interactions, the probability that the system will be scattered into a different state will be small enough, so that to a good approximation these events may be neglected and the system may then be treated as if it were free.

Suppose that  two systems,  $S_{1}$  and  $S_{2}$,  are continuously attracted by each other.  They may then experience continuous interaction events. However, even in such a case it may happen that as a result of the interaction the $S_{2}$ changes its state by only a small enough amount (e.g. if $S_{2}$ is sufficiently heavy), so that that there is only a small probability that it destroys the coherence between the initial and final states of $S_{1}$. \footnote{Of course,  only the coherence between the overlapping parts  of  the in- and outgoing  $S_{1}$ particle states would then be destroyed,  branches of the total state describing the ($S_{1}$  +  $S_{2}$) system would continue to be coherent with each other even in that case (unless the system interacts with and entangles itself with some third object).}  Then, except for those very improbable events,   $S_{2}$ may be considered a "spectator", and to that approximation $S_{1}$ will move as if in a fixed potential. 

Let us suppose that particle A is not free but is acted upon by a force, that this force can not be derived form a fixed potential but is caused by the presence of particle B, and that the interaction of the two particles is sufficiently strong, so that the entanglement of the states referring to A and B has to be taken into account.  As an example, one may think of two particles bound together in a state of well defined energy around the origin of the coordiante system.  The positions of the two particles are then correlated, the energy eigenstate is an entangled state, and now the  $\psi_{j}\varphi_{k} $  of Eq.(\ref{1})  are the correlated position states. Not being energy eigenstes of the system, however, the  $\psi_{j}\varphi_{k} $  continue to evolve, as a result of their continuing interactions the two particles keep on encountering events, therefore reality can continue to move from one $\psi_ {j} \varphi_{k}$  to another in a manner that can not be predicted, but with probabilities that can be calculated. Whichever way reality moves though, it will remain associated with a $\psi_ {j} \varphi_{k}$ branch of the energy eigenstate in question, so that in the absence of any other interaction, the reality content of the energy eigenstate itself will remain unchanged. One can check that this result is consistent with what we would obtain if we did not know, or did not consider the fact that the system under discussion is in fact made up of two constituent particles. In that case one would conclude that in the absence of any other interaction, the system as a whole does not encounter any events, therefore its reality content must indeed remain unchanged.

When the particles are macroscopic, then the time evolution of the system can  be more easily visualized.  This is, because different correlated states of A and B will then interact with the surrounding universe in different ways, therefore these states will soon decohere, until each particle can be described by a well localized wave packet. \footnote{Decoherence will soon occur, but not instantly. For a short time, therefore, a more complicated state can be realized.} According to the CI, only one of these states survives, all others disappear as the wave function "collapses". According to the standard version of the RSI, all such states survive and each is equally real in its own parallel universe. On the other hand, according to the modified version, all such states do indeed survive, but if one particular state is being observed to exist, then that state is real in this universe, all other states are not real, and there are no parallel universes. Therefore, when A and B are macroscopic, reality largely follows the classical motion. Nevertheless,  strictly speaking even in this case, on a sufficiently refined scale, entanglement of A and B persists (for example between their positions in their common rest frame), and only one branch of the entangled state is realized at any one time. When this fact is checked by experiment,  the system will of course be found in only one of those correlated states, from which it follows that then that particular state is the realized one. 

As the interaction between particles increases in strength, it makes less sense to talk about individual constituent  particles, and eventually one may want to consider only the evolution of the entire compound system, in which reality moves in some, possibly hard to visualize manner. This is similar to what one encounters when describing particle collisions: The process of the actual transformation of particles into each other can be much harder to visualize then, say, the behavior of asymptotically free particles in the final state.

\section{ Concluding remarks} 

In the foregoing discussion we repeatedly referred to the concept of  physical ``reality". Much has been said and written over the years in an attempt to understand this concept. Some of its attributes have been clarified, yet it remains true that over the last few thousand years no one was able to give a universally accepted definition of it. For the modest purposes of this discussion, the following working hypothesis/definition may be of some use: ``If you and I are communicating with each other, then both of us are real."  \footnote{To be precise, I should add: "as seen by me". Although not needed here, it is natural to also postulate the reflexive property of reality: If an entity  $A$  is real as seen by me, then I am real as seen by  $A$. The transitive property of reality then follows: If  $A$  is real as seen by  $B$, and  $B$  is real as seen by me, then  $A$  is also real as seen by me.} Here ``you" may be anything, including a mechanical instrument, or some object responding to experimental probing. ``Communicating" is understood to mean sending and receiving messages. Please do not ask how do I know that the messages I receive from you actually do originate with you, or that you --- and maybe everything else --- is not just a figment of my imagination. (I am doing what I can, without allowing ourselves to get hopelessly bogged down.) 
\footnote{An object whose presence is not detected by any observer, might as well not exist as far as the rest of the Universe is concerned.  Detection by an observer means that the observer changes state as a result of the detection process. The fact that it changes, implies that the observer has a choice between at least two distinct outcomes.  It is therefore not surprising that as son as entanglement appears, the concept of ``reality" has to be dealt with.}

Comparing with each other the CI and the two versions of the RSI, one can conclude that (1.) the modified version RSI shares with the standard ``many worlds'' version the property that probability amplitudes do not collapse, but evolve continuously, and (2.) it shares with the CI the property that it is not strictly deterministic, and has built into it exactly the same amount of indeterminacy as does the CI.

The information content of the early universe was determined by the laws of nature and the initial conditions. The universe as we know it today contains vastly more information than that. According to the standard version, this excess information is due to the particular sequence of events that took place on ÒourÓ branch of probability amplitudes: the branch that we can observe. \cite{Hs} However, if one considered todayÕs universe taken as a whole, i.e. taking all branches of all probability amplitudes, it would appear as a rather simple object by comparison: it contains no more information than did the early universe. By contrast, the modified version implies that the sequence of real events represents original information {\it not} determined by the laws of physics and the initial conditions. This additional information can not be predicted, and keeps on growing continuously with time. It is in fact growing at an increasingly fast rate, as all physical systems participate in interactions, certain event outcomes become realized out of the increasingly many possibilities, and states continue to decohere. According to this view, therefore, the universe today (which is the same as the universe taken as a whole, since there are no other branches of reality) is enormously more rich in information than was the early universe, and continues to get richer. To the extent that one may consider information as being part of the physical world, one can therefore say that according to the modified version, the coming into being of the universe did not stop after the Big Bang, but has been going on ever since. 

In the standard interpretation reality is distributed over a vast number of outcomes. Each possible outcome can be considered to have reality value proportional to its probability, a value lying within the continuous range from 0 to 1. By contrast, according to the modified interpretation, only one outcome is real, others are not. Consequently only one amplitude, the one that is realized, has reality value unity, for all others it is zero, no matter what the probabilities may be. In this sense one could say that the modified version predicts that reality content is ``quantized''. It is also ``conserved'' in the sense that reality value can not appear, or disappear, and it may be visualized as being attached to the representative point of the universe as it proceeds through Hilbert space without interruption from one realized event outcome to the next, along any allowed event sequence, but only one sequence at a time. During any event not directly probed, the reality value can not be ascertained. But whenever the next observation occurs, the value is found to be unchanged. \footnote{In this respect it behaves like a conserved net unit charge that moves across a Feynman diagram from the initial to the final state, never disappearing, attaching itself to one particle or another, but not distributing itself over several, while all particles, including the uncharged ones, contribute to the outcome.} All constituent systems belonging to a realized physical system must themselves be realized, so that if one chooses to use this terminology, one may say that the product of reality values belonging to all constituent systems equals the reality value of the total system. Similarly, the reality value of a state equals the sum of the reality values of all its branches.
             
Generally all amplitudes evolve in the same way in both the standard and the modified interpretation (except possibly for some untested cases mentioned below). Therefore all results concerning the behavior of amplitudes can be taken over directly form the standard to the modified interpretation, including insights gained concerning the connection between the quantum and classical worlds, the nature of coherence, and the mechanisms for decoherence.

To predict the probabilities of various possible outcomes of an experiment, in general one needs to know the {\it global}  probability amplitudes. Those can be calculated for example from the laws of physics and the initial conditions of the universe. The initial conditions of the universe, however, are not known. This will usually not cause difficulties, because apart from very special cases, one expects that experimental outcomes can be calculated using only the {\it local} initial conditions together with the laws of physics. The exceptional cases are those in which unexpected other branches of outcomes can interfere with the outcome of the experiment in question. That would happen for example, if the phases of various amplitudes that one would normally expect not to be coherent, would for some reason remain coherent, or would become coherent again, i.e. if they were constrained in special, and as yet unknown ways. This may conceivably occur in a big cosmological crunch. (In such a case local observers would presumably see entropy decreasing, records of the past becoming unstable, and records of the future becoming more permanent. They may then naturally start referring to what we call the future as the past, and vice versa.) In those cases the local initial conditions would have to be supplemented by at least some additional effects of the global probability amplitudes.  These effects would then represent additional input with the compelling force of a natural law, that must be used together with the usual natural laws and local initial conditions to make predictions. In this sense one may consider that type of additional input itself as a natural law, valid at that time and place. 
       
Evidently all experimental predictions of the modified version agree exactly with those that follow from the standard version, except in some esoteric cases. Those could be expected only if the universe behaved differently from what we are now familiar with, e.g. during the very early (or very late) universe, or if certain unexpected features of the natural laws were found to hold. In that case one might be able accumulate evidence in favor of either the standard or the modified interpretation. For example, if it turned out that contrary to all expectations gravitation was not fully quantized, so that all quantum mechanical branches could affect the same physical spacetime, then under certain assumptions the standard version would predict that effects of alternative realities should be detectable in our world, while according to the modified version no alternative realities exist, and therefore no such effects could appear.  Other opportunities to experimentally differentiate the two versions could arise, if quantum mechanics were nonlinear in the sense suggested by S. Weinberg \cite{SW}. (See Appendix.) 

We conclude that the modified version of the Relative State Interpretation is just as self-consistent, and as complete as the standard version.  Both versions lead to precisely the same experimental predictions, except perhaps in rather extreme, as yet untested  cases. As we now understand the structure of Quantum Mechanics, it allows more than one interpretation, in particular it admits both RSI versions.
Today therefore the acceptance of several realities and many worlds is optional, rather than compulsory. The two basic alternatives are:  many worlds, many realities and complete determinism, as required by the standard version; or one world, one reality, and some indeterminacy, as implied by the modified version. At this time the choice between these appears to be a matter of personal preference.

\section*{Acknowledgements}

I am grateful to  Michael Kellman, Robert Zimmermann and especially Steven Hsu for valuable discussions.

\vspace{-.5cm}
\section*{Appendix}  

Assume that in a certain world the time evolution of quantum mechanics is, as proposed by S. Weinberg \cite{SW}, not strictly linear. It was pointed out by N. Gisin \cite{NG}, and by J. Polchinski \cite{JP}  that in such a world supraluminal communication could be physically realized, if the CI  of quantum mechanics is valid.  However \cite{JP}, if the standard version of the RSI holds, then by appropriately restricting the form of nonlinear interactions, such communication can be excluded, but in that case communication between different branches of the wave function would become possible. Inter-branch communication, in turn, can be avoided by requiring that nonlinear interactions have a certain form in some preferred coordinate frame. The theory being nonlinear, results depend on the orientation of axes, but how such a frame should be defined, remains unclear. If one adopts the modified version, communication between real and non-real branches can be excluded by requiring that all terms in the Hamiltonian function \cite{SW} contain products of either only realized amplitudes describing the system under consideration, or only non-realized ones. That defines one preferred axis of the coordinate frame in question, and in the modified version, but not in the standard one, that suffices. (This prescription has to be supplemented if communication between non-real branches were also to be excluded, but that effect could be observable only when originally real and non-real branches interfered.) Whereas in linear quantum mechanics this condition is automatically satisfied, in the nonlinear case it represents an independent postulate. If the system is composed of constituents, and its state contains several branches, only one of them is realized. The Hamiltonian then depends on which that one is. By observing the time evolution of non-linear contributions to  quantum mechanics, one could gain information about which is the realized state, thus "reality" would assume additional significance. To date no evidence has emerged in support of nonlinear quantum mechanics, therefore  delving into more detail here would be premature.

\end{document}